\documentclass[aip,jcp,reprint,nobibnotes,groupedaddress]{revtex4-1}
\usepackage{graphicx}% Include figure files
\usepackage{dcolumn}% Align table columns on decimal point
\usepackage{amsmath}
\newcommand{\bfr}{{\bf r}}
\newcommand{\bfR}{{\bf R}}
\newcommand{\ud}{\mathrm{d}}

\begin{document}

\title{Nonlocal van der Waals density functional: The simpler the better}
\author{Oleg A. Vydrov}
\author{Troy Van Voorhis}
\affiliation{Department of Chemistry, Massachusetts Institute of Technology,\\
 Cambridge, MA, 02139, USA}

\date{\today}

\begin{abstract}
We devise a nonlocal correlation energy functional that describes the entire range of
dispersion interactions in a seamless fashion using only the electron density as input.
The new functional is considerably simpler than its predecessors of a similar type.
The functional has a tractable and robust analytic form that lends itself to efficient
self-consistent implementation. When paired with an appropriate exchange functional,
our nonlocal correlation model yields accurate interaction energies of weakly-bound
complexes, not only near the energy minima but also far from equilibrium. Our model
exhibits an outstanding precision at predicting equilibrium intermonomer separations
in van der Waals complexes. It also gives accurate covalent bond lengths and
atomization energies. Hence the functional proposed in this work is a computationally
inexpensive electronic structure tool of broad applicability.
\end{abstract}
\maketitle

\section{Introduction}

In Kohn-Sham density functional theory (DFT),\cite{Kohn,Primer} only one contribution to
the ground state energy is not known exactly --- the correlation energy. Common
approximations to the correlation energy have the form of local or semilocal density
functionals.\cite{Staroverov} One of the numerous limitations of (semi)local functionals
is their inherent inability to describe such long-range correlation effects as dispersion
(van der Waals) interactions.\cite{ChemRev} Empirical dispersion corrections are quite popular
and reasonably successful, but they typically entail a departure from pure DFT into the realm
of classical force fields.\cite{Grimme} The proper physics of long-range van der Waals
interactions can be captured with fully nonlocal correlation functionals.\cite{ChemRev} However,
the rigor usually comes at the cost of an explicit and cumbersome dependence on Kohn-Sham
orbitals, both occupied and virtual, as exemplified by the random-phase approximation and
related methods (see e.g.\ Refs.~\onlinecite{Nguyen} and \onlinecite{Angyan} and references
therein). An elegant compromise between rigor and computational tractability is achieved
in the recently introduced class of nonlocal correlation functionals that treat the
entire range of dispersion interactions in a general and seamless fashion, yet include no
explicit orbital dependence and use only the electron density as input.\cite{vdW-DF-04,
vdW-DF-10,vdW-DF-09,VV09} In this article we devise a nonlocal van der Waals functional
that is considerably simpler yet more accurate than any of its precursors.\cite{vdW-DF-04,
vdW-DF-10,vdW-DF-09,VV09} We give no derivation for the new functional form, but we show
that it recovers all the relevant limits and satisfies many exact constraints. We suggest
a suitable combination of exchange and correlation terms that performs remarkably well in
all our benchmark tests.

\section{Formalism}

We write the nonlocal correlation energy as
\begin{equation}
 E_\text{c}^\text{nl} = \frac{\hbar}{2} \iint \ud\bfr \ud\bfr'
 \, n(\bfr) \, \Phi(\bfr,\bfr') \, n(\bfr'),
 \label{Enl}
\end{equation}
where $n$ is the total electron density. Building upon the insights gained in
Refs.~\onlinecite{vdW-DF-09,VV09,Comment,*Reply,VV-C6}, we write the
correlation kernel as
\begin{equation}
 \Phi = -\frac{3 \, e^4}{2 m^2 g g' (g + g')}
 \label{Phi}
\end{equation}
with
\begin{equation}
 g = \omega_0(\bfr) R^2 + \kappa(\bfr)
 \label{g}
\end{equation}
and similarly
\begin{equation}
 g' = \omega_0(\bfr') R^2 + \kappa(\bfr').
 \label{g1}
\end{equation}
In the above equations, $R = \left|\bfr-\bfr'\right|$ and
\begin{equation}
 \omega_0(\bfr) = \sqrt{\omega_g^2(\bfr) + \frac{\omega_p^2(\bfr)}{3}},
 \label{w0}
\end{equation}
with the local plasma frequency defined via $\omega_p^2 = 4\pi n e^2\!/m$
and the local band gap given by
\begin{equation}
 \omega_g^2(\bfr) = C \frac{\hbar^2}{m^2}
 \left| \frac{\nabla n(\bfr)}{n(\bfr)} \right|^4,
 \label{gap}
\end{equation}
where $C$ is a parameter adjusted to give accurate asymptotic van der Waals
$C_6$ coefficients, as elaborated in Section~\ref{sec:adjust}.

In the $R \to \infty$ limit,
\begin{equation}
 \Phi \to -\frac{3\,e^4}{2 m^2 \omega_0(\bfr) \omega_0(\bfr') \big[\omega_0(\bfr) + \omega_0(\bfr')\big] R^6},
\end{equation}
so that $E_\text{c}^\text{nl}$ of Eq.~(\ref{Enl}) has precisely the same long-range behavior as the nonlocal
energy in VV09 of Ref.~\onlinecite{VV09}. A detailed analysis and physical justification of this asymptotic
form was given in Ref.~\onlinecite{VV-C6}.

For $R \to 0$, the correlation kernel of Eq.~(\ref{Phi}) behaves as
\begin{equation}
 \Phi = -A + B \, R^2 + \dots,
\end{equation}
in accordance with the result of Koide \cite{Koide} for the $R \to 0$ asymptotic
behavior of the dispersion interaction energy.

In Eqs.~(\ref{g}) and (\ref{g1}), we introduced the quantity
\begin{equation}
 \kappa(\bfr) = b \, \frac{v_F^2(\bfr)}{\omega_p(\bfr)}
 = 3 \, b \, \frac{\omega_p(\bfr)}{k_s^2(\bfr)},
 \label{kappa}
\end{equation}
where $v_F = (3\pi^2 n)^{1/3} \hbar/m$ is the local Fermi velocity,
$k_s = \sqrt{3} \, \omega_p / v_F$ is the Thomas--Fermi screening wave vector,
and $b$ is an adjustable parameter that controls the short-range damping of the
$R^{-6}$ asymptote.

In the uniform density limit, Eq.~(\ref{Phi}) reduces to
\begin{equation}
 \Phi^\text{uni} = -\frac{3 e^4}{4 m^2} \left[ \frac{\omega_p}{\sqrt{3}} \, R^2
 + b \, \frac{v_F^2}{\omega_p} \right]^{-3},
 \label{Phi-1}
\end{equation}
and Eq.~(\ref{Enl}) gives the following energy density per electron:
\begin{align}
 \varepsilon_\text{c}^\text{uni} &= 2\pi\hbar n \!\int_0^\infty\! R^2 \Phi^\text{uni} \,\ud R
  = -\frac{3\pi^2\hbar e^4 n}{32\, m^2 v_F^3} \bigg[\frac{3}{b^2}\bigg]^{3/4} \nonumber \\
 &= -\frac{e^2}{32\, a_0} \bigg[\frac{3}{b^2}\bigg]^{3/4} = -\beta,
 \label{uni}
\end{align}
where $a_0 = \hbar^2 / m e^2$ is the Bohr radius and $\beta$ is a density-independent constant.

It is instructive to rewrite Eq.~(\ref{Phi-1}) in a different form:
\begin{equation}
 \Phi^\text{uni} = -\frac{9\sqrt{3}\, e^4}{4 m^2 \omega_p^3 R^6}
 \left[ 1 + \frac{3\sqrt{3}\, b}{(k_s R)^2} \right]^{-3}.
 \label{Phi-2}
\end{equation}
The above equation shows that the $R^{-6}$ asymptote is damped at short range on the
length scale given by $k_s R$.

Finally, we define our van der Waals density functional as
\begin{align}
 E_\text{c}^\text{VV10} &= E_\text{c}^\text{nl} + \beta N \nonumber \\
 &= \int\!\ud\bfr\, n(\bfr) \bigg[ \beta
  + \frac{\hbar}{2} \int\!\ud\bfr' \, n(\bfr') \, \Phi(\bfr,\bfr') \bigg],
 \label{VV10}
\end{align}
where $N = \int \ud\bfr \, n(\bfr)$ is the number of electrons and $\beta$ is determined
by Eq.~(\ref{uni}). By construction, $E_\text{c}^\text{VV10}$ of Eq.~(\ref{VV10})
vanishes in the uniform density limit. This is a useful property: we can pair
$E_\text{c}^\text{VV10}$ with an existing exchange-correlation (XC) functional without
affecting the description of the uniform electron gas.

It is a nontrivial task to determine how Eq.~(\ref{VV10}) behaves in the slowly varying density
limit. At any rate, it has been argued that recovery of the density gradient expansion for
correlation energy is of little relevance for real systems.\cite{Perdew-06}

The van der Waals functional of Eq.~(\ref{VV10}) is very simple and easily implementable.
The double integral in Eq.~(\ref{Enl}) is in practice evaluated as a double sum over a
numerical grid. For $R \to 0$, the kernel of Eq.~(\ref{Phi}) goes to a constant, avoiding
any complications with the numerical integration. In the inner integration loop, only simple
arithmetic operations need to be performed. Hence this new functional is computationally less
expensive than the nonlocal functionals of Refs.~\onlinecite{vdW-DF-04,vdW-DF-10,vdW-DF-09,VV09}.

The previous version of the nonlocal van der Waals functional, VV09 proposed by us in
Ref.~\onlinecite{VV09}, was more computationally demanding because it used a rather
elaborate damping function. Evaluation of the nonlocal correlation energy in VV09 required
computing a square root, an exponent, and an error function for each pair of grid points
in the double sum. The energy derivatives, needed for the self-consistent implementation
of VV09, were also quite complicated.\cite{VV09-imp} With the new VV10 model of Eq.~(\ref{VV10}),
these energy derivatives are much simplified. The self-consistent implementation of VV10 within
a Gaussian basis set code is described in detail in the next section.

\section{Implementation}

For the sake of brevity, in this section we switch to atomic units ($\hbar=m=e=1$).
In these units, $\kappa$ of Eq.~(\ref{kappa}) is expressed rather simply as
\begin{equation}
 \kappa(\bfr) = b \, \frac{3\pi}{2}\! \left[\frac{n(\bfr)}{9 \pi}\right]^{1/6}.
\end{equation}

Our implementation of VV10 is very similar to the implementation of its
predecessor VV09, described in Ref.~\onlinecite{VV09-imp}.
We express the electron density via a Gaussian basis set:
\begin{equation}
 n(\bfr) = \sum_{\mu\nu} P_{\mu\nu}\, \chi_\mu(\bfr) \chi_\nu(\bfr),
\end{equation}
where $\chi_\mu$ and $\chi_\nu$ are basis functions, and $P_{\mu\nu}$ are the density
matrix elements. For the self-consistent treatment, we need to find the derivatives of
$E_\text{c}^\text{VV10}$ with respect to $P_{\mu\nu}$:
\begin{equation}
 \frac{\ud E_\text{c}^\text{VV10}}{\ud P_{\mu\nu}} =
 \int\! \ud\bfr \, \chi_\mu(\bfr)
 \frac{\delta E_\text{c}^\text{VV10}}{\delta n(\bfr)} \chi_\nu(\bfr).
 \label{mat-el}
\end{equation}
To that end, we employ the standard formalism\cite{Pople-JCP-93} developed for
semilocal XC functionals:
\begin{equation}
 \frac{\ud E_\text{c}^\text{VV10}}{\ud P_{\mu\nu}} =
 \int\! \ud\bfr \Big[ F_n \chi_\mu \chi_\nu
 + 2 F_\gamma \nabla n \cdot \nabla \big( \chi_\mu \chi_\nu \big) \Big].
 \label{Fc}
\end{equation}
We denote $\gamma = |\nabla n|^2$ for convenience.
$F_n$ and $F_\gamma$ in Eq.~(\ref{Fc}) are computed as
\begin{align}
 F_n(\bfr) = & \; \beta + \int\! \ud\bfr' \, n(\bfr') \, \Phi \nonumber \\
 & + n(\bfr) \bigg[ \frac{\partial\kappa}{\partial n}(\bfr) U(\bfr)
 + \frac{\partial\omega_0}{\partial n}(\bfr) W(\bfr) \bigg]
\end{align}
and
\begin{equation}
 F_\gamma(\bfr) = n(\bfr) \frac{\partial\omega_0}{\partial \gamma}(\bfr) W(\bfr),
\end{equation}
where
\begin{equation}
 U(\bfr) = - \!\int\! \ud\bfr'\, n(\bfr') \,
 \Phi \bigg[\frac{1}{g} + \frac{1}{g+g'}\bigg]
\end{equation}
and
\begin{equation}
 W(\bfr) = - \!\int\! \ud\bfr'\, n(\bfr') \left|\bfr-\bfr'\right|^2
 \Phi \bigg[\frac{1}{g} + \frac{1}{g+g'}\bigg].
\end{equation}
In the above equations, $\Phi$, $g$, and $g'$ are assumed to be functions of both
$\bfr$ and $\bfr'$.

Within an atom-centered basis set implementation, the gradient of $E_\text{c}^\text{VV10}$
with respect to the displacement of nucleus $A$ has three contributions:
\begin{equation}
 \nabla_A E_\text{c}^\text{VV10} = {\bf g}^A_\text{GBF} + {\bf g}^A_\text{weights}
 + {\bf g}^A_\text{grid}.
 \label{grad}
\end{equation}

${\bf g}^A_\text{GBF}$ denotes the contribution of the Gaussian basis functions.
This term can be evaluated by plugging $F_n$ and $F_\gamma$ into Eq.~(9) of
Ref.~\onlinecite{Pople-JCP-93} instead of $\partial f/ \partial n$ and
$\partial f/ \partial \gamma$.

The last two terms in Eq.~(\ref{grad}) are due to the use of atom-centered numerical
integration quadratures. We employ the atomic partitioning scheme of Becke,\cite{Becke-W}
which separates the molecular integral into atomic contributions:
\begin{eqnarray}
 E_\text{c}^\text{VV10} &=&  \sum_{A} \sum_{i \in A} w_{Ai} \, n(\bfr_{Ai})
 \bigg[ \beta  \nonumber \\
 &+& \frac{1}{2} \sum_{B} \sum_{j \in B} w_{Bj} \, n(\bfr_{Bj})
 \, \Phi(\bfr_{Ai},\bfr_{Bj}) \bigg],
\end{eqnarray} 
where $w_{Ai}$ and $w_{Bj}$ are the quadrature weights, and the grid points
$\bfr_{Ai}$ are given by $\bfr_{Ai} = \bfR_A + \bfr_i$, where $\bfR_A$ is the
position of nucleus $A$, with the $\bfr_i$ defining a one-center integration grid.
The quadrature weights depend on the nuclear configuration and hence have a nonzero
gradient with respect to nuclear displacements:
\begin{eqnarray}
 {\bf g}^A_\text{weights} &=& \sum_{B} \sum_{i \in B} \big( \nabla_A w_{Bi} \big)
 n(\bfr_{Bi}) \bigg[ \beta \nonumber \\
 &+& \sum_{C} \sum_{j \in C} w_{Cj} \, n(\bfr_{Cj})
 \, \Phi(\bfr_{Bi},\bfr_{Cj}) \bigg].
\end{eqnarray}
The weight derivatives $\nabla_A w_{Bi}$ can be found in Ref.~\onlinecite{Pople-JCP-93}.

The last term in Eq.~(\ref{grad}) arises because each one-center quadrature grid
moves together with its parent nucleus and the nonlocal correlation kernel $\Phi$
depends explicitly on the distance between the grid points.
The ${\bf g}^A_\text{grid}$ term can be computed as:
\begin{align}
 {\bf g}^A_\text{grid} =& \sum_{i \in A} w_{Ai} \, n(\bfr_{Ai}) \nonumber \\
 \times& \sum_{B \ne A} \sum_{j \in B} w_{Bj} \, n(\bfr_{Bj})
 Q(\bfr_{Ai},\bfr_{Bj}) \, \big(\bfr_{Ai}-\bfr_{Bj}\big),
\end{align}
where
\begin{equation}
 Q(\bfr,\bfr') = -2 \Phi \left[ \frac{\omega_0(\bfr)}{g} + \frac{\omega_0(\bfr')}{g'}
 + \frac{\omega_0(\bfr)+\omega_0(\bfr')}{g+g'} \right].
\end{equation}

Analytic gradients, computed via Eq.~(\ref{grad}), enable us to perform structural
optimizations routinely and efficiently.

\section{Adjustments
\label{sec:adjust}}

$E_\text{c}^\text{VV10}$ of Eq.~(\ref{VV10}) can in principle be paired with nearly any existing
XC functional. However, to avoid double-counting as much as possible, it is preferable to
combine $E_\text{c}^\text{VV10}$ with a functional that gives no significant binding in van der
Waals complexes. Since predictions of XC functionals for weakly interacting systems can differ
drastically, the parameter $b$, controlling the short-range behavior of $E_\text{c}^\text{VV10}$,
has to be adjusted for a particular parent XC approximation.

The Perdew--Wang 86 (PW86) exchange functional\cite{PW86} has been shown to
describe the repulsive parts of van der Waals potentials rather well,\cite{Lacks,Becke-09,*Becke-10,
Langreth-JCTC-09} and a refitted version of PW86 was recently proposed.\cite{Langreth-JCTC-09}
We denote this `refitted PW86' as rPW86 here. In the rest of this article we will consider one
particularly apt combination of the exchange and correlation terms:
\begin{equation}
 E_\text{xc}^\text{VV10} = E_\text{x}^\text{rPW86} + E_\text{c}^\text{PBE} + E_\text{c}^\text{VV10},
 \label{XC}
\end{equation}
where $E_\text{c}^\text{PBE}$ is the semilocal correlation energy functional in the generalized
gradient approximation of Perdew, Burke, and Ernzerhof (PBE).\cite{PBE} Hereafter, we will often
refer to the XC functional of Eq.~(\ref{XC}) simply as VV10.

There are two adjustable parameters in our method, but only $C$ in Eq.~(\ref{gap}) affects the
asymptotic dispersion $C_6$ coefficients. We fit $C$ to minimize the average error for the
benchmark set of 54 $C_6$ coefficients compiled in Tables II and III of Ref.~\onlinecite{VV-C6}.
Using self-consistent electron densities from rPW86-PBE (i.e. using $E_\text{xc}
= E_\text{x}^\text{rPW86} + E_\text{c}^\text{PBE}$), we find the optimal value of $C = 0.0093$,
which gives the mean absolute percentage error (MAPE) of 14\% for the whole test set of 54
species. The largest errors in $C_6$ occur for metal atoms. When all the metal atoms
(Li, Na, Mg, Al, Zn, Ga) are removed from the benchmark set and only the remaining 48
species are considered, the MAPE in $C_6$ drops to 9\%. Note that a slightly smaller value of the
parameter $C$ was used in Refs.~\onlinecite{VV09}, \onlinecite{VV-C6}, and \onlinecite{VV09-imp},
since the source of electron densities was different.\cite{C-note}

Another adjustable parameter $b$, introduced in Eq.~(\ref{kappa}), controls the short range
damping of $\Phi$. Using $E_\text{xc}^\text{VV10}$ of Eq.~(\ref{XC}), we fitted $b$ on the training
set of 22 binding energies (the S22 set of Ref.~\onlinecite{S22}), with the computational details
described in Section~\ref{sec:details} and the results given in Section~\ref{sec:S22}. We obtained
the best fit with $b=5.9$. Correspondingly, in Eq.~(\ref{VV10}) we use
\[
 \beta = \frac{1}{32} \bigg[\frac{3}{5.9^2}\bigg]^{3/4} = 0.00497
\]
in the Hartree atomic units.

Although the rPW86 exchange functional appears to be a very good match for our correlation model,
this choice is certainly not unique. In the Appendix we describe an alternative choice of exchange
which is nearly equally suitable for this purpose.

\section{Computational details
\label{sec:details}}

\begin{table}[bp!]
\caption{Binding energies (in kcal/mol) for the S22 test set. All calculations were performed
 self-consistently with the aug-cc-pVTZ basis set and counterpoise corrected. Molecular
 structures are from Ref.~\onlinecite{S22} and reference binding energies are from Ref.~\onlinecite{S22A}
 for all systems except Adenine--Thymine complexes, for which the numbers from Ref.~\onlinecite{S22B}
 were used.
\label{Table:S22}}
\begin{ruledtabular}
\begin{tabular}{lddd}
Complex (symmetry) & \multicolumn{1}{c}{Ref.} & \multicolumn{1}{c}{VV10} & \multicolumn{1}{c}{vdW-DF2} \\ \hline
\multicolumn{4}{l}{\hspace{4mm}\emph{Dispersion-bound complexes} (8)}\\
CH$_4$ dimer ($D_\text{3d}$)			& 0.53  & 0.50  & 0.68 \\
C$_2$H$_4$ dimer ($D_\text{2d}$)		& 1.48  & 1.42  & 1.32 \\
Benzene--CH$_4$ ($C_3$)				& 1.45  & 1.45  & 1.29 \\
Benzene dimer ($C_\text{2h}$)\footnotemark[1]	& 2.66  & 2.71  & 2.15 \\
Pyrazine dimer ($C_\text{s}$)			& 4.26  & 4.02  & 3.30 \\
Uracil dimer ($C_2$)\footnotemark[2]		& 9.78  & 9.70  & 8.76 \\
Indole--Benzene ($C_1$)\footnotemark[2]		& 4.52  & 4.54  & 3.44 \\
Adenine--Thymine ($C_1$)\footnotemark[2]	& 11.66 & 11.42 & 9.58 \\
\multicolumn{4}{l}{\hspace{4mm}\emph{Mixed complexes} (7)} \\
C$_2$H$_4$--C$_2$H$_2$ ($C_\text{2v}$)		& 1.50  & 1.68  & 1.53 \\
Benzene--H$_2$O ($C_\text{s}$)			& 3.28  & 3.31  & 2.80 \\
Benzene--NH$_3$ ($C_\text{s}$)			& 2.32  & 2.28  & 1.99 \\
Benzene--HCN ($C_\text{s}$)			& 4.54  & 4.30  & 3.55 \\
Benzene dimer ($C_\text{2v}$)\footnotemark[3]	& 2.72  & 2.54  & 2.06 \\
Indole--Benzene ($C_\text{s}$)\footnotemark[3]	& 5.63  & 5.27  & 4.20 \\
Phenol dimer ($C_1$)				& 7.10  & 6.99  & 5.97 \\
\multicolumn{4}{l}{\hspace{4mm}\emph{Hydrogen-bonded complexes} (7)}\\
NH$_3$ dimer ($C_\text{2h}$)			& 3.15  & 3.43  & 2.97  \\
H$_2$O dimer ($C_\text{s}$)			& 5.00  & 5.50  & 4.78  \\
Formic acid dimer ($C_\text{2h}$)		& 18.75 & 19.96 & 16.77 \\
Formamide dimer ($C_\text{2h}$)			& 16.06 & 16.71 & 14.43 \\
Uracil dimer ($C_\text{2h}$)\footnotemark[4]	& 20.64 & 21.10 & 18.69 \\
2-pyridone--2-aminopyridine ($C_1$)		& 16.94 & 18.05 & 15.37 \\
Adenine--Thymine WC ($C_1$)\footnotemark[4]	& 16.74 & 17.42 & 14.74 \\
\end{tabular}
\end{ruledtabular}
\footnotetext[1]{`Parallel-displaced' configuration.}
\footnotetext[2]{Stacked configuration.}
\footnotetext[3]{T-shaped configuration.}
\footnotetext[4]{Planar configuration.}
\end{table}

VV10 has been implemented self-consistently into the \textsc{Q-Chem} software package.\cite{Q-Chem}
In our assessment of VV10, we compare its performance to another recent van der Waals
functional of a similar type --- vdW-DF2 of Ref.~\onlinecite{vdW-DF-10}. Since vdW-DF2 is just a
reparameterization of an earlier functional, vdW-DF of Ref.~\onlinecite{vdW-DF-04}, we use the same
implementation\cite{Vydrov-08} for both of them. The full XC energy in vdW-DF2 is defined as
\begin{equation}
 E_\text{xc}^\text{vdW-DF2} = E_\text{x}^\text{rPW86} + E_\text{c}^\text{LDA}
 + E_\text{c-nl}^\text{vdW-DF2},
\end{equation}
where $E_\text{c}^\text{LDA}$ is the correlation energy in the local density approximation,
for which we use the parameterization of Perdew and Wang.\cite{PW92}

All reported calculations were performed self-consistently. For weakly bound systems, the interaction
energies were counterpoise corrected. The unpruned Euler--Maclaurin--Lebedev (99,590) quadrature grid
was used to evaluate all local and semilocal contributions ($E_\text{x}^\text{rPW86}$,
$E_\text{c}^\text{PBE}$, and $E_\text{c}^\text{LDA}$) and the (75,302) grid was used for the nonlocal
components. Deviations of computed binding energies from the reference values are analyzed
with the help of mean errors (ME), mean absolute errors (MAE), and mean absolute percentage errors
(MAPE). In all tables, binding energies are reported as positive values. Hence a negative ME
indicates underbinding while a positive ME means overbinding.

In our recent article\cite{VV09-imp} assessing the performance of VV09, we used the same benchmark
sets of weakly bound systems and the same computational details. Therefore, the results reported in
Ref.~\onlinecite{VV09-imp} can be directly compared to the results obtained with the new functionals,
presented in Section~\ref{sec:weak} below.

\section{Results for weakly interacting systems
\label{sec:weak}}
\subsection{The S22 benchmark set
\label{sec:S22}}

\begin{table}[bp!]
\caption{Summary of deviations from the reference values of the binding energies reported in
Table~\ref{Table:S22}. ME and MAE are in kcal/mol, MAPE is in percents.
\label{Table:Errors}}
\begin{ruledtabular}
\begin{tabular}{@{\extracolsep{0.2\columnwidth}} ldd}
& \multicolumn{1}{c}{VV10} & \multicolumn{1}{c}{vdW-DF2} \\ \hline
\multicolumn{3}{l}{\hspace{6mm}\emph{Dispersion-bound complexes} (8)}\\
ME   & -0.07 & -0.73 \\
MAE  &  0.09 &  0.76 \\
MAPE (\%) &  2.6  &  18.0 \\
\multicolumn{3}{l}{\hspace{6mm}\emph{Mixed complexes} (7)} \\
ME   & -0.10 & -0.71 \\
MAE  &  0.16 &  0.72 \\
MAPE (\%) &  4.8  &  16.8 \\
\multicolumn{3}{l}{\hspace{6mm}\emph{Hydrogen-bonded complexes} (7)}\\
ME   &  0.70 & -1.36 \\
MAE  &  0.70 &  1.36 \\
MAPE (\%) &  6.1  &  8.8  \\
\multicolumn{3}{l}{\hspace{6mm}\emph{Total} (22)}\\
ME   &  0.16 & -0.92 \\
MAE  &  0.31 &  0.94 \\
MAPE (\%) &  4.4  &  14.7 \\
\end{tabular}
\end{ruledtabular}
\end{table}

\begin{figure*}[tbp!]
\includegraphics[width=0.95\textwidth]{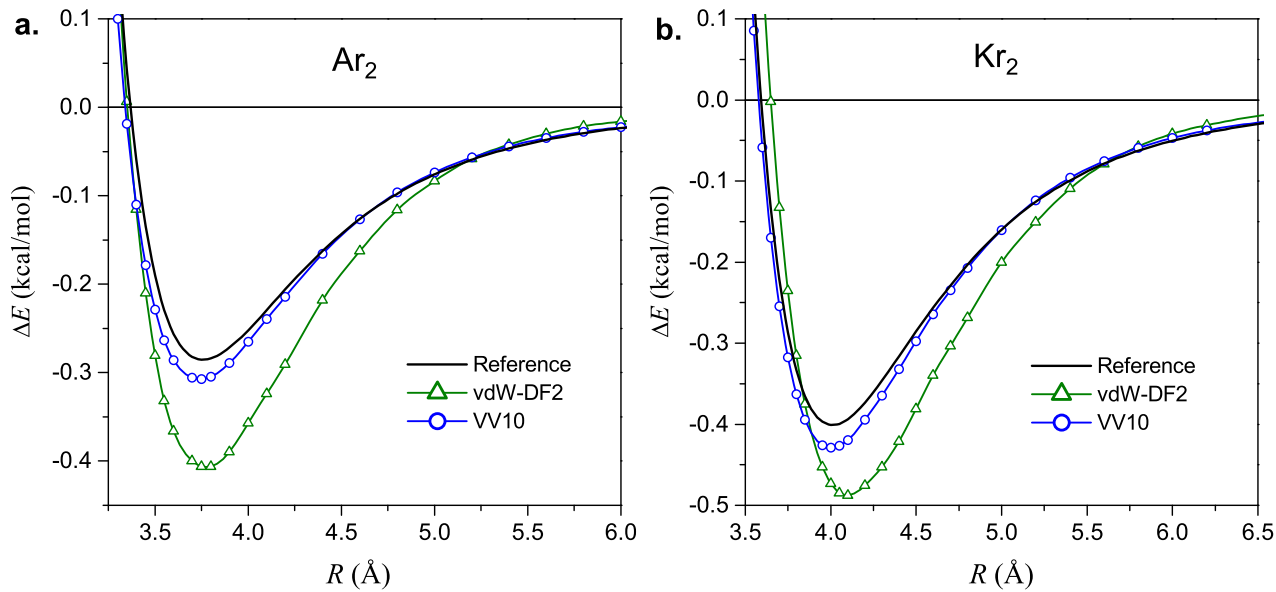}
\caption{Interaction energy curves for (a) the argon dimer and (b) the krypton dimer.
 Reference curves are from Ref.~\onlinecite{Tang-03}. VV10 and vdW-DF2 results were
 obtained self-consistently with the aug-cc-pVQZ basis set.}
 \label{fig1}
\end{figure*}

Binding energies for the S22 test set, computed with vdW-DF2 and VV10 at the geometries of
Ref.~\onlinecite{S22}, are given in Table~\ref{Table:S22}. The error statistics are
summarized in Table~\ref{Table:Errors}. Updated reference values of binding
energies for the S22 set were recently obtained in Refs.~\onlinecite{S22A} and \onlinecite{S22B}.
For our benchmark set, we picked the values that we deemed the most converged with respect to
the basis set size: we took the reference values for most of the systems from Ref.~\onlinecite{S22A},
except for the Adenine--Thymine complexes (in both stacked and planar configurations), for which
we used the numbers from Ref.~\onlinecite{S22B}.

As Tables~\ref{Table:S22} and \ref{Table:Errors} show, VV10 performs remarkably well for all
dispersion-bound and mixed complexes, but yields somewhat larger errors for hydrogen-bonded
systems. The parameter $b$ inside $E_\text{c}^\text{VV10}$ was fitted to the binding energies
of the S22 set, but it is by no means trivial to achieve such accuracy by adjusting just one
global parameter.

vdW-DF2 overbinds some of the smaller systems, specifically (CH$_4$)$_2$ and C$_2$H$_4$--C$_2$H$_2$,
but underbinds all other complexes in the S22 set, when the geometries from Ref.~\onlinecite{S22} are
used. This underbinding tendency intensifies as the monomer sizes increase. A negative ME for the
binding energies of the S22 set was also found in Ref.~\onlinecite{vdW-DF-10}. However, intermonomer
separations were optimized in Ref.~\onlinecite{vdW-DF-10}, yielding larger binding energies, as
compared to our results for the fixed\cite{S22} geometries.

Binding energies calculated for a standard set of near-equilibrium geometries certainly do not tell
the full story about the performance of a functional. For a more comprehensive assessment, it is
necessary to determine whether equilibrium intermonomer separations are accurately located and
whether reasonable interaction energies are predicted far from equilibrium. To this end, we analyze
binding energy curves for several complexes.

\begin{figure*}[tbp!]
\includegraphics[width=0.96\textwidth]{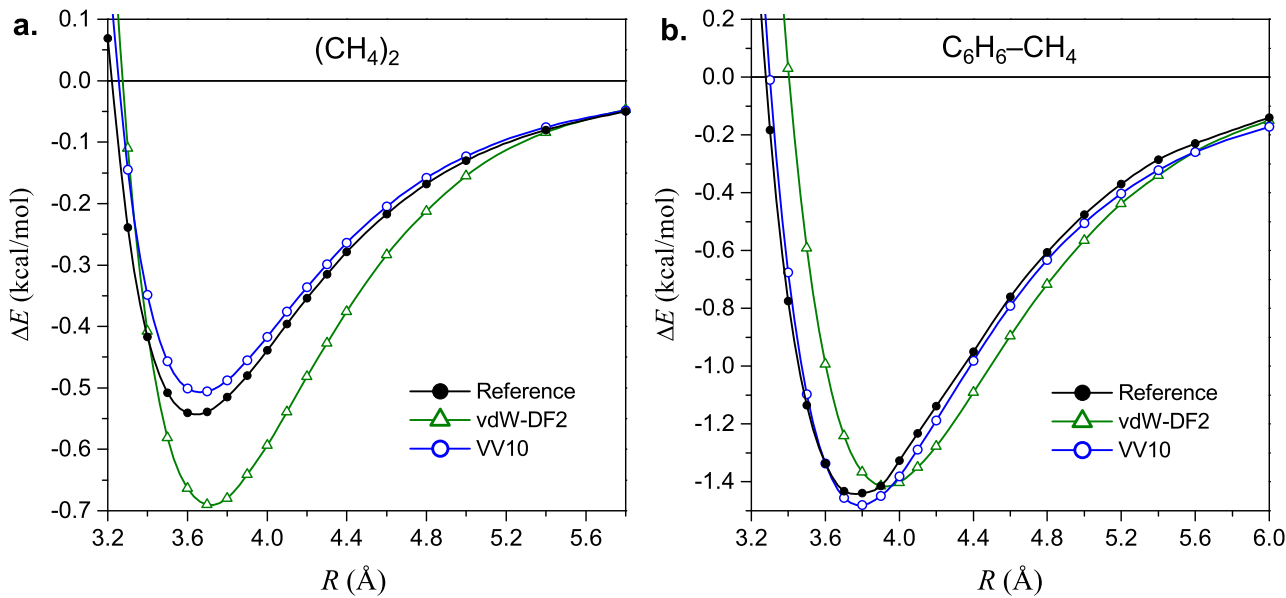}
\caption{Interaction energy curves for (a) the methane dimer and (b) the benzene--methane complex.
 $R$ is the distance between the centers of mass of the monomers. Reference values are from
 Ref.~\onlinecite{Sherrill-09}. VV10 and vdW-DF2 results were obtained self-consistently with the
 aug-cc-pVTZ basis set.}
 \label{fig2}
\end{figure*}

\begin{figure*}[tbp!]
\includegraphics[width=0.96\textwidth]{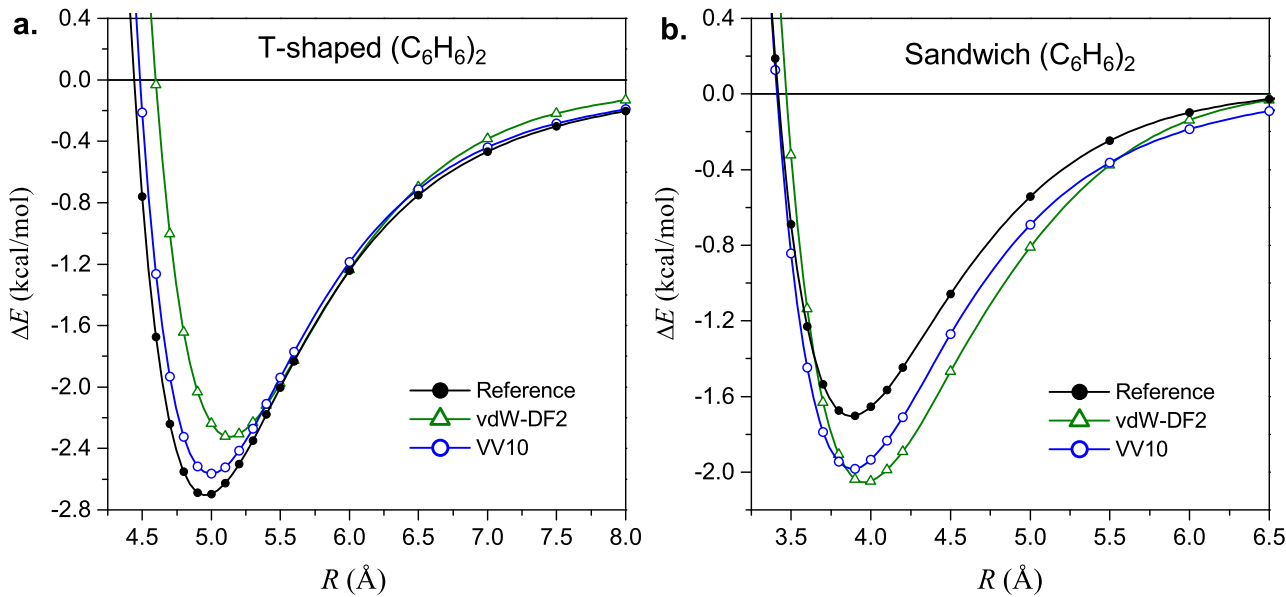}
\caption{Interaction energy curves for (a) the T-shaped and (b) the stacked sandwich-shaped
 benzene dimers. $R$ is the distance between the centers of mass of the monomers. Reference values
 are from Ref.~\onlinecite{Sherrill-09}. VV10 and vdW-DF2 results were obtained self-consistently
 with the aug-cc-pVTZ basis set.}
 \label{fig3}
\end{figure*}

\subsection{Interaction energy curves}

We computed interaction energy curves for six weakly-bound systems using VV10 and vdW-DF2.
Figure~\ref{fig1} shows the results for the argon and krypton dimers, Figure~\ref{fig2} for
the methane dimer ($D_\text{3d}$ symmetry) and the benzene--methane complex ($C_\text{3v}$),
and Figure~\ref{fig3} for benzene dimers in two configurations: T-shaped ($C_\text{2v}$) and
stacked sandwich-shaped ($D_\text{6h}$). We applied counterpoise corrections at all
intermonomer distances. For Ar$_2$ and Kr$_2$, the reference values were taken from
Ref.~\onlinecite{Tang-03}, and for the other complexes from Ref.~\onlinecite{Sherrill-09}.
For molecular complexes, we used the same fixed monomer geometries as in
Ref.~\onlinecite{Sherrill-09}.

For all six complexes, VV10 reproduces the equilibrium intermonomer separations with
remarkable precision. Interaction energies yielded by VV10 agree quite well with
the reference values at all separations. The well depths are predicted with reasonable
accuracy: VV10 slightly underbinds (CH$_4$)$_2$ and the T-shaped (C$_6$H$_6$)$_2$, but it
slightly overbinds Ar$_2$ and Kr$_2$ and more strongly overbinds the sandwich-shaped
(C$_6$H$_6$)$_2$. For the benzene--methane complex, the deviations of VV10 from the
reference curve are within the uncertainties of the reference\cite{Sherrill-09}
interaction energies at nearly all separations. The long-range falloff of the potential
energy curves is reproduced very well by VV10 in nearly all cases.

vdW-DF2 tends to predict the energy minima at somewhat too large separations.
It significantly overbinds the complexes of small monomers: Ar$_2$, Kr$_2$, and
(CH$_4$)$_2$. For all systems in Figures~\ref{fig1}, \ref{fig2}, and \ref{fig3},
except Ar$_2$, the repulsive walls given by vdW-DF2 are too steep and shifted
towards larger distances as compared to the reference curves. As was shown in
Ref.~\onlinecite{VV-C6}, vdW-DF2 strongly underestimates asymptotic $C_6$
coefficients. As a result, for systems where the asymptotic interactions are
dominated by dispersion, vdW-DF2 will yield too shallow potential energy curves at
long range. This underestimation is visible at large separations for Ar$_2$, Kr$_2$,
and the T-shaped (C$_6$H$_6$)$_2$. The effects of the poor vdW-DF2 asymptotics are
expected to be more noticeable at very large intermonomer distances or for very
large monomers. Such cases are not significantly represented in this study.

As Figure~\ref{fig3}.(b) shows, both VV10 and vdW-DF2 significantly overestimate the
equilibrium binding energy in the sandwich-shaped benzene dimer. This may be partially
caused by the lack of Axilrod-Teller-type three-body interactions in the functionals of
the form of Eq.~(\ref{Enl}). Another possible source of errors is the rPW86 exchange
functional. It was found in Ref.~\onlinecite{Langreth-JCTC-09} that PW86 exchange is more
repulsive than Hartree--Fock for the T-shaped benzene dimer, but less repulsive for the
sandwich-shaped configuration.

\section{Results for atoms and covalent bonds}

We have shown that VV10 describes weakly-bound systems quite well. In this section,
we show that VV10 is a broadly applicable method that also treats covalent and
ionic bonds well.

\subsection{Total energies of atoms}

\begin{table}[tbp!]
\caption{Total energies of atoms in Hartree atomic units. The exact values are from
 Refs.~\onlinecite{atomic-E1} and \onlinecite{atomic-E2}.
\label{Table:atoms}}
\begin{ruledtabular}
\begin{tabular}{ldddd}
 & \multicolumn{1}{c}{Exact} & \multicolumn{1}{c}{PBE}
 & \multicolumn{1}{c}{rPW86-PBE} & \multicolumn{1}{c}{VV10} \\ \hline
H & -0.500 & -0.500 & -0.509 & -0.505 \\
He & -2.904 & -2.893 & -2.926 & -2.916 \\
Li & -7.478 & -7.462 & -7.517 & -7.503 \\
Be & -14.667 & -14.630 & -14.709 & -14.692 \\
B & -24.654 & -24.612 & -24.723 & -24.700 \\
C & -37.845 & -37.799 & -37.941 & -37.914 \\
N & -54.589 & -54.536 & -54.708 & -54.677 \\
O & -75.067 & -75.015 & -75.230 & -75.194 \\
F & -99.734 & -99.676 & -99.932 & -99.891 \\
Ne & -128.938 & -128.866 & -129.159 & -129.114 \\
Na & -162.255 & -162.172 & -162.502 & -162.453 \\
Mg & -200.053 & -199.955 & -200.325 & -200.271 \\
Al & -242.346 & -242.236 & -242.645 & -242.588 \\
Si & -289.359 & -289.234 & -289.682 & -289.621 \\
P & -341.259 & -341.115 & -341.602 & -341.536 \\
S & -398.110 & -397.952 & -398.481 & -398.411 \\
Cl & -460.148 & -459.974 & -460.544 & -460.470 \\
Ar & -527.540 & -527.346 & -527.955 & -527.876 \\ \hline
ME/\=e\footnotemark[1] &
   & 0.008 & -0.019 & -0.014 \\
\end{tabular}
\end{ruledtabular}
\footnotetext[1]{Mean error per electron.}
\end{table}

In Table~\ref{Table:atoms} we report the total energies of atoms up to Ar,
computed with rPW86-PBE (that is rPW86 exchange\cite{Langreth-JCTC-09} with
PBE correlation\cite{PBE}) as well as VV10, which is rPW86-PBE with the addition of
$E_\text{c}^\text{VV10}$, as shown in Eq.~(\ref{XC}). We also include the popular
semilocal XC functional PBE.\cite{PBE} We used the aug-pc-4 basis set\cite{aug-pc-4}
with all beyond-$f$ polarization functions removed. This basis set is expected to
yield energies close to the basis set limit. As the benchmark, we use the accurate
nonrelativistic values of atomic energies from Refs.~\onlinecite{atomic-E1} and
\onlinecite{atomic-E2}.

Among the three considered functionals, PBE gives the most accurate atomic
energies. PBE total energies are systematically above the exact values, except
for the H atom, for which PBE gives a nearly exact energy. On the contrary,
rPW86-PBE and VV10 yield total energies that are systematically too low.

We find $E_\text{c}^\text{VV10}$ of Eq.~(\ref{VV10}) to be positive not only for
atoms, but for all systems considered in this study. We have no proof that
$E_\text{c}^\text{VV10}$ is always strictly nonnegative, i.e.\ that $\beta N \ge
-E_\text{c}^\text{nl}$, but we have found no exceptions so far.
The positive contribution of $E_\text{c}^\text{VV10}$ brings the rPW86-PBE
total energies in better agreement with the exact values, as
Table~\ref{Table:atoms} shows.

We did not include vdW-DF2 in Table~\ref{Table:atoms} because this functional is
not defined for open-shell systems. For the five closed-shell atoms in the set of
Table~\ref{Table:atoms}, vdW-DF2 yields the ME of $-0.048$ hartree per electron.
vdW-DF2 total energies are much too low primarily because this functional includes
$E_\text{c}^\text{LDA}$ as a constituent and LDA is known to severely overestimate
the magnitudes of correlation energies in atoms.

\subsection{Atomization energies}

We computed atomization energies for the AE6 set developed by Lynch and Truhlar.
\cite{AE6} This set includes only six molecules, but it is quite diverse and was
constructed to be representative, that is to reproduce mean errors in much larger
sets. We used the molecular geometries suggested by the authors of the
set.\cite{AE6} The results and the mean errors are given in Table~\ref{Table:AE6}.

rPW86-PBE, which is the backbone of VV10, yields considerably more accurate
atomization energies as compared to the popular PBE functional. The addition of
the nonlocal correlation term $E_\text{c}^\text{VV10}$ to rPW86-PBE increases the
atomization energies in all cases, as the VV10 results in Table~\ref{Table:AE6}
show. In other words, the addition of $E_\text{c}^\text{VV10}$ makes covalent
bonds somewhat stronger. As a result, VV10 yields smaller errors than rPW86-PBE
for three molecules in the AE6 set, but larger errors for the other three molecules.
On average, the performance of VV10 for the AE6 set is quite similar to its parent
functional rPW86-PBE.

\begin{table}[tbp!]
\caption{Atomization energies for the AE6 set of Ref.~\onlinecite{AE6}
 computed with the aug-cc-pVTZ basis set. All values are in kcal/mol.
 \label{Table:AE6}}
\begin{ruledtabular}
\begin{tabular}{ldddd}
 & \multicolumn{1}{c}{Ref.} & \multicolumn{1}{c}{PBE}
 & \multicolumn{1}{c}{rPW86-PBE} & \multicolumn{1}{c}{VV10} \\ \hline
S$_2$       &  101.7 &  112.5 &  106.0 &  108.5 \\
SiO         &  192.1 &  193.4 &  190.1 &  191.8 \\
SiH$_4$     &  322.4 &  311.6 &  308.2 &  310.2 \\
glyoxal     &  633.4 &  661.7 &  643.9 &  650.1 \\
propyne     &  704.8 &  720.1 &  706.1 &  711.5 \\
cyclobutane & 1149.0 & 1166.1 & 1137.0 & 1148.8 \\ \hline
ME   &  & 10.4 & -2.0 & 2.9 \\
MAE  &  & 14.0 &  7.4 & 7.2 \\
\end{tabular}
\end{ruledtabular}
\end{table}

\subsection{Bond lengths}

\begin{table*}[tbp!]
\caption{Equilibrium bond lengths $(r_e)$ computed using the aug-cc-pVTZ basis set.
 All values are in \AA. Experimental data are from Ref.~\onlinecite{CRC}.
 \label{Table:bonds}}
\begin{ruledtabular}
\begin{tabular}{lcccccc}
 & Expt. & PBE & rPW86-LDA & vdW-DF2 & rPW86-PBE & VV10 \\ \hline
H$_2$  & 0.741 & 0.751 & 0.734 & 0.736 & 0.745 & 0.745 \\
LiF    & 1.564 & 1.583 & 1.592 & 1.592 & 1.586 & 1.585 \\
LiCl   & 2.021 & 2.028 & 2.041 & 2.041 & 2.031 & 2.031 \\
Li$_2$ & 2.673 & 2.726 & 2.663 & 2.670 & 2.699 & 2.700 \\
CH$_4$ & 1.087 & 1.096 & 1.088 & 1.090 & 1.092 & 1.093 \\
CO     & 1.128 & 1.137 & 1.136 & 1.136 & 1.137 & 1.137 \\
CO$_2$ & 1.160 & 1.172 & 1.174 & 1.173 & 1.172 & 1.172 \\
CS$_2$ & 1.553 & 1.566 & 1.573 & 1.571 & 1.568 & 1.567 \\
N$_2$  & 1.098 & 1.103 & 1.101 & 1.101 & 1.102 & 1.102 \\
HF     & 0.917 & 0.932 & 0.934 & 0.935 & 0.934 & 0.934 \\
F$_2$  & 1.412 & 1.413 & 1.459 & 1.453 & 1.433 & 1.432 \\
NaCl   & 2.361 & 2.385 & 2.410 & 2.401 & 2.390 & 2.385 \\
NaBr   & 2.502 & 2.528 & 2.560 & 2.547 & 2.538 & 2.532 \\
Na$_2$ & 3.079 & 3.079 & 3.006 & 2.987 & 3.035 & 3.024 \\
SiO    & 1.510 & 1.536 & 1.543 & 1.540 & 1.539 & 1.538 \\
P$_2$  & 1.893 & 1.911 & 1.922 & 1.918 & 1.914 & 1.913 \\
HCl    & 1.275 & 1.291 & 1.289 & 1.290 & 1.290 & 1.290 \\
Cl$_2$ & 1.988 & 2.019 & 2.085 & 2.073 & 2.040 & 2.037 \\
HBr    & 1.415 & 1.433 & 1.436 & 1.437 & 1.435 & 1.435 \\
Br$_2$ & 2.281 & 2.311 & 2.396 & 2.380 & 2.343 & 2.336 \\ \hline
ME     &       & 0.017 & 0.024 & 0.021 & 0.018 & 0.017 \\
MAE    &       & 0.017 & 0.033 & 0.031 & 0.023 & 0.022 \\
\end{tabular}
\end{ruledtabular}
\end{table*}

In Table~\ref{Table:bonds} we have compiled a test set of 20 small closed-shell
molecules for which experimental\cite{CRC} equilibrium bond lengths $(r_e)$ are known.
The test set consists mostly of diatomics, but includes several polyatomic molecules
of high symmetry whose geometry is completely determined by a single bond length.
Most of the molecules in the set are covalently bound, except for the alkali metal
halides where the bonding is ionic.

Table~\ref{Table:bonds} reports the results of geometry optimizations
carried out using analytic gradients and the aug-cc-pVTZ basis set. We tested
vdW-DF2 and VV10 alongside their semilocal parent functionals, rPW86-LDA and
rPW86-PBE respectively. We also included PBE for the sake of comparison.
For most molecules in Table~\ref{Table:bonds}, vdW-DF2 yields nearly the same
bond lengths as rPW86-LDA, and VV10 nearly the same as rPW86-PBE. Hence the
addition of a nonlocal correlation term has an almost negligible effect on
covalent and ionic bond lengths. The largest change occurs for Na$_2$: the Na--Na
bond length given by VV10 is 0.011 {\AA} shorter than with rPW86-PBE. This is not
surprising since the Na--Na bond is weaker than most covalent bonds.
On average for the test set of Table~\ref{Table:bonds}, the performance of VV10
is on par with PBE, while vdW-DF2 is somewhat less accurate.

\section{Conclusions}

The VV10 correlation model proposed in this work belongs to the family\cite{vdW-DF-04,vdW-DF-10,
vdW-DF-09,VV09} of nonlocal van der Waals density functionals defined in terms of the
electron density alone and using no orbital input. These functionals\cite{vdW-DF-04,vdW-DF-10,
vdW-DF-09,VV09} are general and seamless: they require neither splitting the system into
interacting fragments nor any kind of atomic partitioning. They treat inter- and
intra-molecular dispersion interactions on equal footing.

VV10 has the same long-range behavior as its precursor VV09, but the damping mechanism of
dispersion interactions at short range is greatly simplified in VV10. This simplification not
only makes the functional more efficient and computationally tractable, but it also leads
to improved overall accuracy.

An essential aspect of the VV10 formalism is the additional flexibility introduced with the
help of an adjustable parameter $b$ which controls the short range behavior of the nonlocal
correlation energy. When $E_\text{c}^\text{VV10}$ of Eq.~(\ref{VV10}) is added as a correction
to an existing XC functional, $b$ is adjusted to attain a balanced
merging of interaction energy contributions at short and intermediate ranges. A particularly
successful functional is constructed by adding $E_\text{c}^\text{VV10}$ with $b=5.9$ to rPW86-PBE,
as shown in Eq.~(\ref{XC}). Another parameter $C = 0.0093$ inside $E_\text{c}^\text{VV10}$ ensures
that accurate asymptotic $C_6$ coefficients are obtained.\cite{C-note}

As our benchmark tests clearly demonstrate, the functional of Eq.~(\ref{XC}) is a broadly
accurate electronic structure tool: its outstanding predictive power for weakly-bound systems
is complemented by its good description of covalent bonds.

\begin{acknowledgments}
This work was supported by an NSF CAREER grant No. CHE-0547877 and a Packard Fellowship.
\end{acknowledgments}

\appendix*
\section{VV10 with long-range corrected exchange}

The XC functional of Eq.~(\ref{XC}) includes the semilocal rPW86 exchange term as a constituent.
Semilocal exchange functionals are known to suffer from self-interaction error (SIE), which
causes such artifacts as poor description of charge transfer complexes and transition states of
chemical reactions. Long-range corrected (LC) hybrid exchange functionals have proven
to be very effective at minimizing the SIE.\cite{LC-wPBE,Vydrov-07} LC hybrids have also been
shown to adequately describe the repulsive parts of van der Waals potentials.\cite{Hirao-07}
We test one particular LC hybrid for its suitability as a counterpart for the VV10 correlation
model --- the LC-$\omega$PBE functional.\cite{LC-wPBE,Henderson-09} The exchange component in
LC-$\omega$PBE is defined as a sum of two parts: the long-range Hartree--Fock (LR-HF) part and
the short-range PBE (SR-PBE) part, for which we use the parameterization of
Ref.~\onlinecite{Henderson-09}. For the range-separation parameter $\omega$, we use the value
of $0.45$~bohr$^{-1}$, as suggested in Ref.~\onlinecite{Henderson-09} (this variant was termed
LC-$\omega$PBE08 therein).
The full XC functional, which we denote LC-VV10, has the form
\begin{equation}
 E_\text{xc}^\text{LC-VV10} = E_\text{x}^\text{SR-PBE} + E_\text{x}^\text{LR-HF}
 + E_\text{c}^\text{PBE} + E_\text{c}^\text{VV10}.
 \label{LC}
\end{equation}
The two parameters inside $E_\text{c}^\text{VV10}$ are adjusted using the same procedure as
described in Section \ref{sec:adjust}. As we previously found,\cite{VV09,VV-C6,C-note}
$C = 0.0089$ gives accurate $C_6$ coefficients at LC-$\omega$PBE electron densities.
We fit the parameter $b$ to the binding energies of the S22 set using the functional
of Eq.~(\ref{LC}) and obtain the optimal value of $b = 6.3$.

In Table \ref{Table:LC} we summarize the errors of LC-VV10 for the test sets used previously
in this article. As compared to the VV10 model with the rPW86 exchange, LC-VV10 performs somewhat
better for binding energies of hydrogen bonded complexes, for total energies of atoms, and for
bond lengths. However, no improvement is observed for atomization energies, while binding
energies of dispersion-bound complexes are slightly worsened on average. Therefore, we suggest
using LC-VV10 only when SIE is an issue. In most other cases the VV10 model of Eq.~(\ref{XC})
is preferable since the semilocal rPW86 exchange is computationally cheaper and much more
easy to implement than an LC hybrid.

\begin{table}[tbp!]
\caption{Errors of LC-VV10 for several test sets. LC-VV10 is defined by Eq.~(\ref{LC}) with
 $\omega = 0.45$, $C = 0.0089$, and $b = 6.3$.
 \label{Table:LC}}
\begin{ruledtabular}
\begin{tabular}{lddd}
 & \multicolumn{1}{c}{ME} & \multicolumn{1}{c}{MAE} & \multicolumn{1}{c}{MAPE (\%)} \\ \hline
\emph{S22 subsets} (kcal/mol) \\
Dispersion-bound & 0.01 & 0.17 & 7.1 \\
Mixed            & 0.09 & 0.16 & 4.0 \\
Hydrogen-bonded  & 0.17 & 0.31 & 2.3 \\
Full S22 set     & 0.09 & 0.21 & 4.6 \\
\emph{Other test sets} \\
Energies of atoms (a.u.) & 0.008\footnotemark[1] & 0.008\footnotemark[1] \\
AE6 (kcal/mol) & 0.5 & 6.9 \\
Bond lengths (\AA) & -0.006 & 0.014 \\
\end{tabular}
\end{ruledtabular}
\footnotetext[1]{Errors per electron.}
\end{table}

It is noteworthy that our model can achieve good performance using a pre-existing and
unmodified exchange functional (rPW86 or LC-$\omega$PBE) and adjusting only the parameters
in the nonlocal correlation. It might be possible to further improve the performance by
tailoring an empirical exchange functional for our specific purpose, in the vein of the
recent works of Refs.~\onlinecite{Pernal,Klimes,Cooper}. However, we prefer to keep
empirical fitting to a minimum.


\begin{thebibliography}{10}
\bibitem{C-note}%
  \BibitemOpen
  \bibinfo {note} {The value of $C = 0.0089$ was obtained in
  Ref.~\onlinecite{VV09} by fitting to a $C_6$ test set using electron
  densities produced by the long-range corrected (LC) hybrid functional
  LC-$\omega$PBE. As compared to LC hybrids, the description of density tails
  by semilocal XC functionals is systematically less accurate, which is
  reflected in $\omega_g$. The readjusted value of $C = 0.0093$ is tailored for
  use with semilocal XC functionals, such as rPW86-PBE.}\BibitemShut{Stop}%
\end{thebibliography}
\end{document}